\documentclass[fleqn,usenatbib]{mnras}

\usepackage{amsmath}
\usepackage{amssymb}	
\usepackage{graphicx}	

\title[Updated spin-down ephemeris for AR Sco]{An Improved Spin-Down Rate for the Proposed White-Dwarf Pulsar AR~Scorpii }

\author[Y. Gaibor et al.]{Y. Gaibor,$^{1,2}$
P. M. Garnavich$^1$,
C. Littlefield$^1$,
S. B. Potter$^{3,4}$,
D. A. H. Buckley$^3$
\\
$^1${Department of Physics, University of Notre Dame, Notre Dame, IN 46556, USA}\\
$^2${Missouri State University, Springfield, MO 65897, USA} \\
$^3${South African Astronomical Observatory, PO Box 9, Observatory 7935, Cape Town, South Africa} \\
$^4${Department of Physics, University of Johannesburg, PO Box 524, Auckland Park 2006, South Africa}
}

\date{Accepted 2020 June 25. Received 2020 June 24; in original form 2020 May 11}

\pubyear{2020}

\begin{document}
\label{firstpage}
\pagerange{\pageref{firstpage}--\pageref{lastpage}}
\maketitle


\begin{abstract}

We analyze rapid-cadence, multiwavelength photometry of AR Scorpii from three observatories, covering five observing seasons. We measure the arrival times of the system's beat pulses and use them to compute an updated ephemeris. The white dwarf spin-down rate is estimated with an uncertainty of only 4\%. These results confirm, beyond any doubt, that the white dwarf's spin period is increasing at the rate consistent with by that of Stiller et al. (2018). We study the evolution of the beat pulse's color index across the orbit. The color of the primary pulse maxima varies significantly across the orbit, with the peaks being bluer after superior conjunction than in the first half of the orbit. Specifically, at orbital phase 0.5, the color index of the primary pulse shows a very sharp discontinuity towards bluer indices. This supports the Potter \&\ Buckley (2018b) synchrotron emission model where the two emitting poles differ significantly in color. However, no corresponding jump in the color of the secondary pulses is seen. Furthermore, our analysis reveals that the arrival times of the pulses can differ by as much as 6~s in simultaneous $u$ and $r$ photometry, depending on the binary orbital phase. If left uncorrected, this wavelength-dependent timing offset could lead to erroneous measurements of the spin-period derivative, particularly with heterogeneous datasets.

\end{abstract}



\section{Introduction}
\label{intro}

AR Scorpii (AR~Sco hereafter) is a binary system consisting of a low-mass red dwarf star and a rapidly spinning white dwarf (WD) with highly unusual characteristics \citep{Marsh}.  The system has a 3.56-hour orbital period, and a spectacular variation in brightness on a two minute time-scale that is related to the spin of the WD and its magnetic interaction with the red dwarf secondary. The pulsed emission is seen over a broad range of wavelengths from radio to soft X-rays \citep{Marsh, takata18, marcote17,stanway, stiller18}. One of the most unusal features of AR~Sco is that the pulsed emission does not appear to be due to accretion on to the WD \citep{Marsh}, as is seen in intermediate polar-type cataclysmic variable star (CVs). The strongly linearly polarized pulses are consistent with synchrotron radiation coming from near the WD \cite{Buckley}, although where and how the electrons are accelerated remains uncertain. \citet{garnavich19} detected slingshot prominences originating from the red dwarf star that implies a magnetic field of several hundred Gauss. This suggests the possibility that magnetic reconnection events between the WD and secondary star fields are the source of the energetic electrons. 

\citet{Marsh} and \citet{stiller18} detected a secular decrease in the period of the pulses that imply that the WD spin rate is decreasing and that the dynamical energy loss is sufficient to explain the excess luminosity seen from the system. Although the WD spin periods of some intermediate polars have been observed to vary on timescales of years \citep{patterson}, the underlying physical mechanisms are very different than those in AR~Sco. Spin-period derivatives in accreting intermediate polars are widely understood to be caused by the competition between a spin-up torque from accretion and a spin-down torque from the WD's magnetic field lines dragging through the accretion flow. In contrast, AR~Sco has been called a ``WD pulsar'' because, just as in neutron-star pulsars, its pulsed, non-thermal electromagnetic emission is exceptionally periodic and appears to be powered by the spin-down of a compact stellar component.

Here, we combine observations of AR~Sco obtained from several sources over five seasons to provide tight constraints on the spin-down rate  of the WD and improve the pulse ephemerides. Further, we search for color dependence in the pulse timings and color variations in the pulses over an orbit.

\begin{figure*}
	\includegraphics[width=\textwidth]{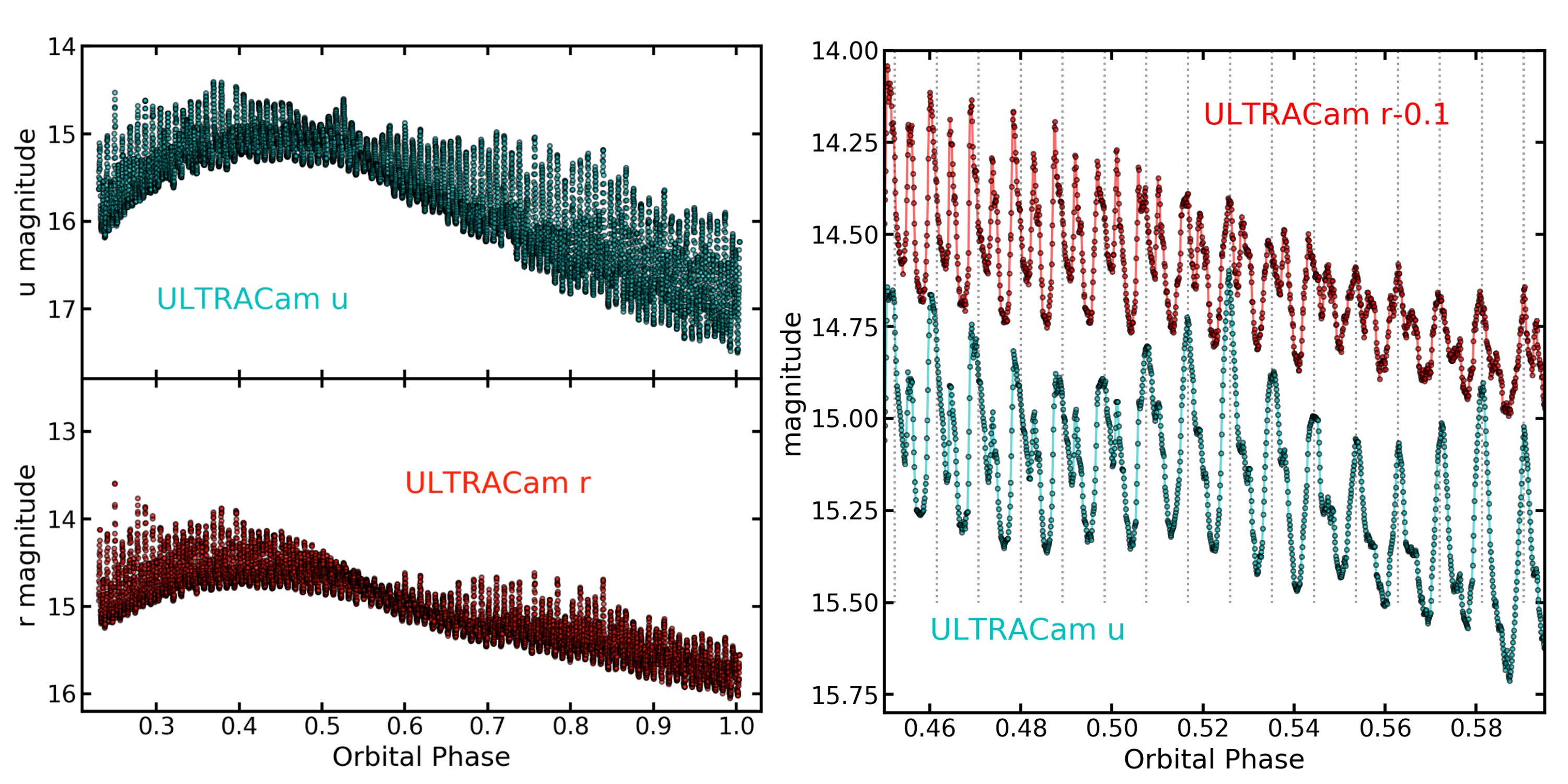}
	\caption{{\bf left:} The ULTRACam light curves from 2015 for
	the $SDSS-r$ and $SDSS-u$ filters. The red and blue light curves show significant differences starting after orbital phase 0.5. For example, the pulses in the $u$-band are stronger than in $r$. Note the extremely blue pulses around phase 0.52 that are quite weak in the $r$~filter. {\bf right:} A close-up of the $u$ and $r$ light curves around stellar conjunction. The secondary pulses nearly disappear in the UV while they are comparable to the primary pulses in the red. The vertical dotted lines mark the interval of the beat period.
	\label{lightcurve}}
\end{figure*}

\section{Data}

The rapid variability of the AR~Sco pulses means that it is imperative to sample the light curve with fast cadence observations.
It is also critical for the time-stamps of individual observations to be accurate. For these reasons we have restricted the sources of data to three observatories: South African Astronomical Observatory (SAAO), the William Herschel Telescope (WHT), and the Sarah L. Krizmanich Telescope (SLKT) on the University of Notre Dame campus.
As noted by \citet{stiller18}, the shutter lag of the camera used on the SLKT was consistent from night-to-night, but its value was uncertain on the level of $\sim$1~s, so these data are analyzed separately from the other sets.

Times for all the observations have been converted to Barycentric Julian Day (BJD) in Barycentric Dynamical Time to remove the Earth's motion and light-travel time effects \citep[for a detailed discussion of these corrections, see][]{eastman}.

\subsection{SLKT}

We analyze photometry data from the 80-cm Sarah L. Krizmanich Telescope (SLKT) obtained over four seasons from 2016 to 2019. \citet{stiller18} published results from the SLKT data taken from 2016, 2017, and 2018, and here we add new observations from the 2019 season. A total of 40.3 hours of data was obtained from 35 observing runs with a typical cadence of 5 to 6 sec. The data was taken unfiltered which corresponded to an effective central wavelength similar to the Johnson-Cousins~$R$ band.

\subsection{SAAO}

We analyze photometry data from the South African Astronomical Observatory (SAAO) using the high-speed photomultiplier instrument \citep[HIPPO;][]{potter08}.  Observations of AR~Sco analyzed here were obtained from 2016 to 2019 with a typical cadence of 2~sec for a total data set covering 17.3 hours. The data was taken unfilter or with an OG570 longpass filter using a red-sensitive photomultiplier detector. Some of these observations have been published in \citet{potter18a}. This photometry had a number of noise glitches that were identified and removed. This was done by running a 5-point median filter over the data and deleting individual observations that exceeded 15$\sigma$ from the distribution of deviations about the median. 

\subsection{WHT}
We also used the WHT ULTRACam photometry of 2015 \citep{Marsh}, which was taken with the r, g, and u filters. This observing run covered 80\%\ of an orbit, with typical cadence of ~1.3 sec. For our analysis, we used the $r$-band data to best match the red color of the SAAO data. We refer to the combined SAAO and WHT $r$-band datasets as SAAO+WHT-r.

\section{Analysis}

\subsection{The Beat Ephemeris}

Power spectra of the light curve of AR~Sco are dominated by a frequency at $\nu_{spin} =$8.5282~mHz attributed to the spin of the WD and a frequency at $\nu_{beat} =$8.4603~mHz identified as the ``beat'' between the WD spin and the orbital period \citep{Marsh, potter18b, stiller18} \footnote{$\nu_{beat} = \nu_{spin} - \nu_{orb}$}. The beat-pulse profile is conspicuously double-peaked, with a strong primary pulse and a weaker secondary pulse during each 118.2~s (1.97~min) beat cycle. A particularly excellent example of this structure is shown in the WHT/ULTRACAM data in Figure~\ref{lightcurve}. 

\citet{stiller18} showed that the most precise method of estimating the change in spin period is to measure the timing of individual pulse peaks over a long baseline.  To find the brightness peaks, we iteratively create a window of 55~s around the predicted peaks, based on the \cite{stiller18} ephemeris. Then, we fit a Gaussian function to the points to estimate the precise time of the peak. We find the corresponding orbital phase by using the \cite{Marsh} orbital ephemeris. The resulting beat-pulse timings and corresponding uncertainties are shown in Table~\ref{beat_SAAO} (SAAO+WHT-r) and in Table~\ref{beat_SLKT} (SLKT). The quality of the averaged timings depend on several factors, including the number of beat pulses measured in the run and their orbital phases. There are about 108 primary beat pulses per binary orbit, so the SAAO+WHT observations cover all or most of an orbit, but the SLKT observations tend to have captured half an orbit or less.

\begin{figure}
   \centering
	\includegraphics[width=1.1\columnwidth]{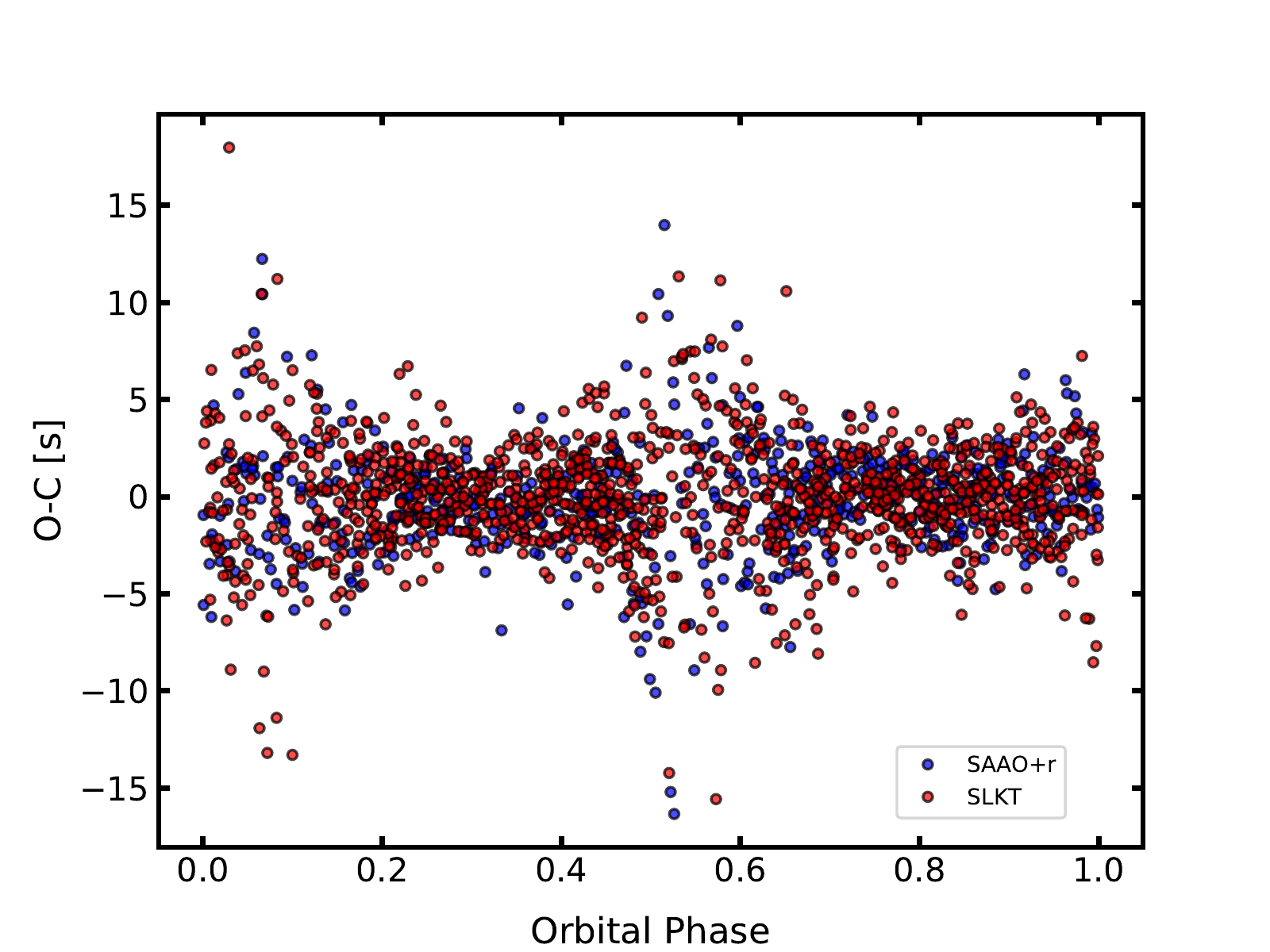}
	\caption{O-C after correcting for orbital dependent phase variations. The SAAO+WHT-r result are shown as blue points and the SLKT points are red. The dispersion in $O-C$ clearly increases around orbital phases 0.05 and 0.55. These larger uncertainties are used to de-weight pulses occurring in these orbital phase ranges. 
	\label{orb_correction}}
\end{figure}

Timings of the primary beat pulse show that its arrival times vary systematically over an orbital cycle \citep{takata17, stiller18}. \citet{stiller18} successfully modeled the primary pulse's phase shifts as the interference between the beat and spin frequencies across the orbit, although other causes are possible. To correct for orbital-phase dependence of the pulse arrival times, we computed a Fourier-series representation of the pulse O$-$C (observed-calculated)\footnote{For an overview of O$-$C analysis, see \citet{sterken}.} as a function of orbital phase, used it to calculate the systematic offset at the orbital phase of the observation, and subtracted this value from the O$-$C of each pulse, as was done in \citet{stiller18}. This process was applied separately to the SAAO+WHT-r data and the SKLT data (Figure~\ref{orb_correction}).

\begin{table}
	\caption{Beat-pulse timings for SAAO+WHT-r \label{beat_SAAO}}
	\begin{tabular}{cccc}
	\hline
	{UT} & {T$_{max}$ BJD} & {Epoch} & {N$_{beat}$$^{a}$} \\
	\hline	
	2015-06-24 & 2457198.383725(51) & $-$543319 & 77 \\
    2016-05-27 & 2457536.416814(38) & $-$296227 & 187 \\
    2017-05-17 & 2457891.415095(46) & $-$36734  & 86  \\
    2018-03-22 & 2458200.518220(43) & 189211  & 110 \\
    2019-05-09 & 2458613.445166(45) & 491048  & 135 \\
    \hline
    \end{tabular}
    $^{a}${Number of peaks measured}
\end{table}

\begin{table}
	\caption{Beat-pulse timings for SLKT \label{beat_SLKT}}
	\begin{tabular}{cccc}
	\hline
	{UT} &  {T$_{max}$ BJD} & {Epoch} & N$_{beat}$$^{a}$ \\
	\hline
    2016-07-28 & 2457597.620408(14) & $-$251489 & 31 \\
    2016-08-03 & 2457603.593295(60) & $-$247123 & 38 \\
    2016-08-22 & 2457622.570821(17) & $-$233251 & 25 \\
    2016-08-23 & 2457623.558578(81) & $-$232529 & 31 \\
    2016-09-01 & 2457632.557571(11) & $-$225951 & 19 \\
    2016-09-02 & 2457633.558975(23) & $-$225219 & 22 \\
    2016-09-03 & 2457634.549437(11) & $-$224495 & 16 \\
    2016-09-04 & 2457635.544029(11) & $-$223768 & 15 \\
    2017-04-23 & 2457866.786138(41) & $-$54737  & 67 \\
    2017-05-07 & 2457880.742939(53) & $-$44535  & 52 \\
    2017-05-08 & 2457881.760770(94) & $-$43791  & 35 \\
    2017-05-15 & 2457888.680346(10) & $-$38733  & 25 \\
    2017-05-17 & 2457890.690013(87) & $-$37264  & 37 \\
    2017-06-01 & 2457905.823344(54) & $-$26202  & 61 \\
    2017-06-02 & 2457906.715299(46) & $-$25550  & 42 \\
    2017-06-03 & 2457907.668816(90) & $-$24853  & 15 \\
    2017-07-07 & 2457941.623732(94) & $-$33     & 39 \\
    2017-08-12 & 2457977.599235(14) & 26264   & 21 \\
    2018-02-26 & 2458175.923470(45) & 171233  & 25 \\
    2018-03-18 & 2458195.917428(17) & 185848  & 20 \\
    2018-03-25 & 2458202.925993(12) & 190971  & 19 \\
    2018-03-26 & 2458203.934218(71) & 191708  & 65 \\
    2018-04-18 & 2458226.809307(83) & 208429  & 27 \\
    2018-04-21 & 2458229.836786(88) & 210642  & 13 \\
    2018-05-24 & 2458262.753367(61) & 234703  & 58 \\
    2018-07-01 & 2458300.649583(99) & 262404  & 39 \\
    2019-03-23 & 2458565.838503(22) & 456249  & 33 \\
    2019-04-13 & 2458586.861255(11) & 471616  & 58 \\
    2019-04-27 & 2458600.820818(73) & 481820  & 47 \\
    2019-05-05 & 2458608.841664(51) & 487683  & 80 \\
    2019-05-06 & 2458609.733619(14) & 488335  & 11 \\
    2019-06-08 & 2458642.613278(60) & 512369  & 40 \\
    2019-06-11 & 2458645.677713(52) & 514609  & 28 \\
    2019-06-14 & 2458648.649092(14) & 516781  & 27 \\
    2019-07-12 & 2458676.650228(18) & 537249  & 27 \\
    \hline
    \end{tabular}
    $^{a}${Number of peaks measured}
\end{table}

\begin{figure*}
    \centering
	\includegraphics[width=0.8\textwidth]{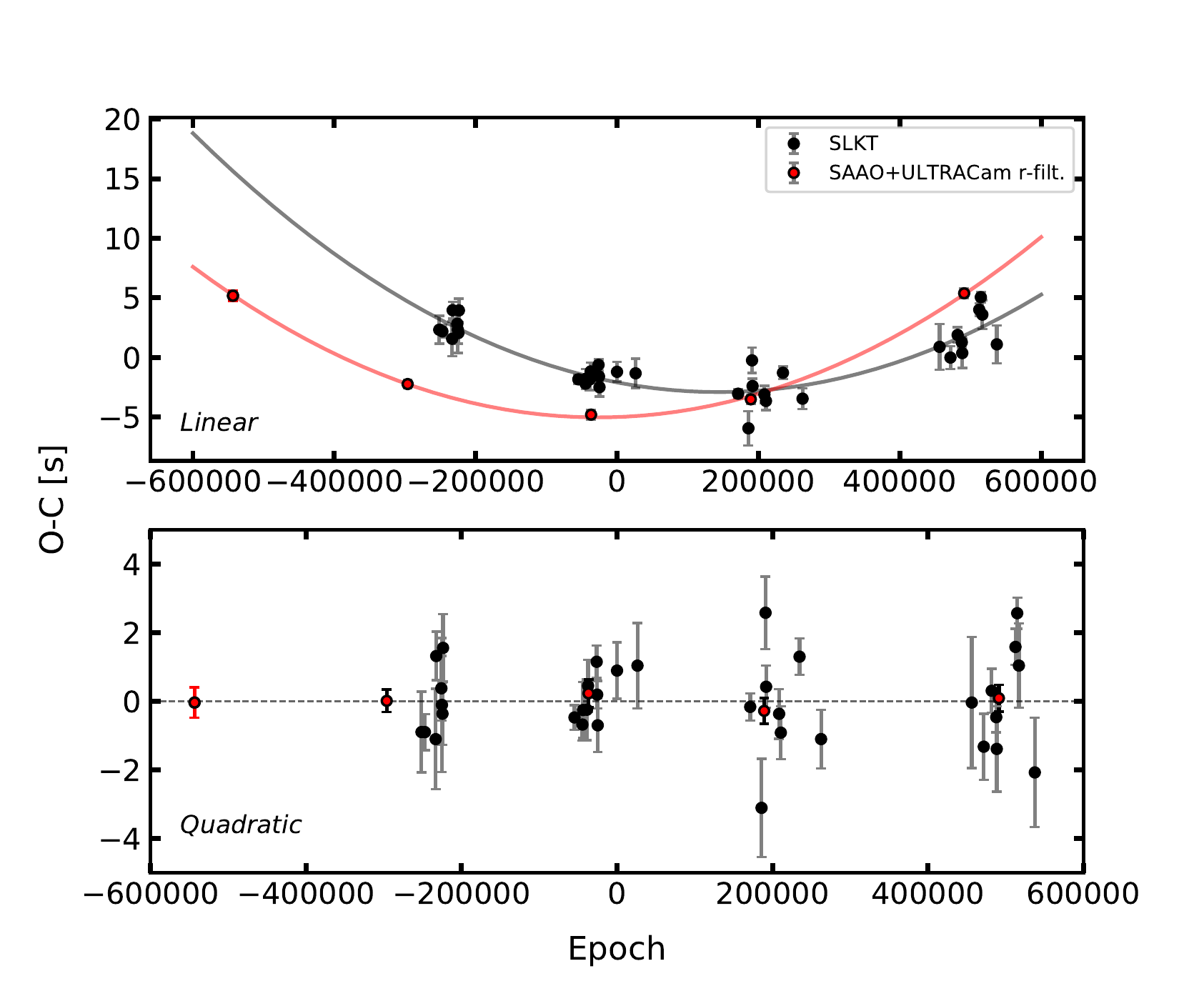}
	\caption{O-C comparing the preliminary polynomial fits of the SLKT (black markers) and SAAO+WHT-r (red markers) data. {\bf Top:}Residuals of the linear fit. {\bf Bottom:}Residuals of a second order polynomial fit. It is clear that a quadratic fit will better describe the data. \label{residuals}}
\end{figure*}
\begin{figure*}
    \centering
	\includegraphics[width=\textwidth]{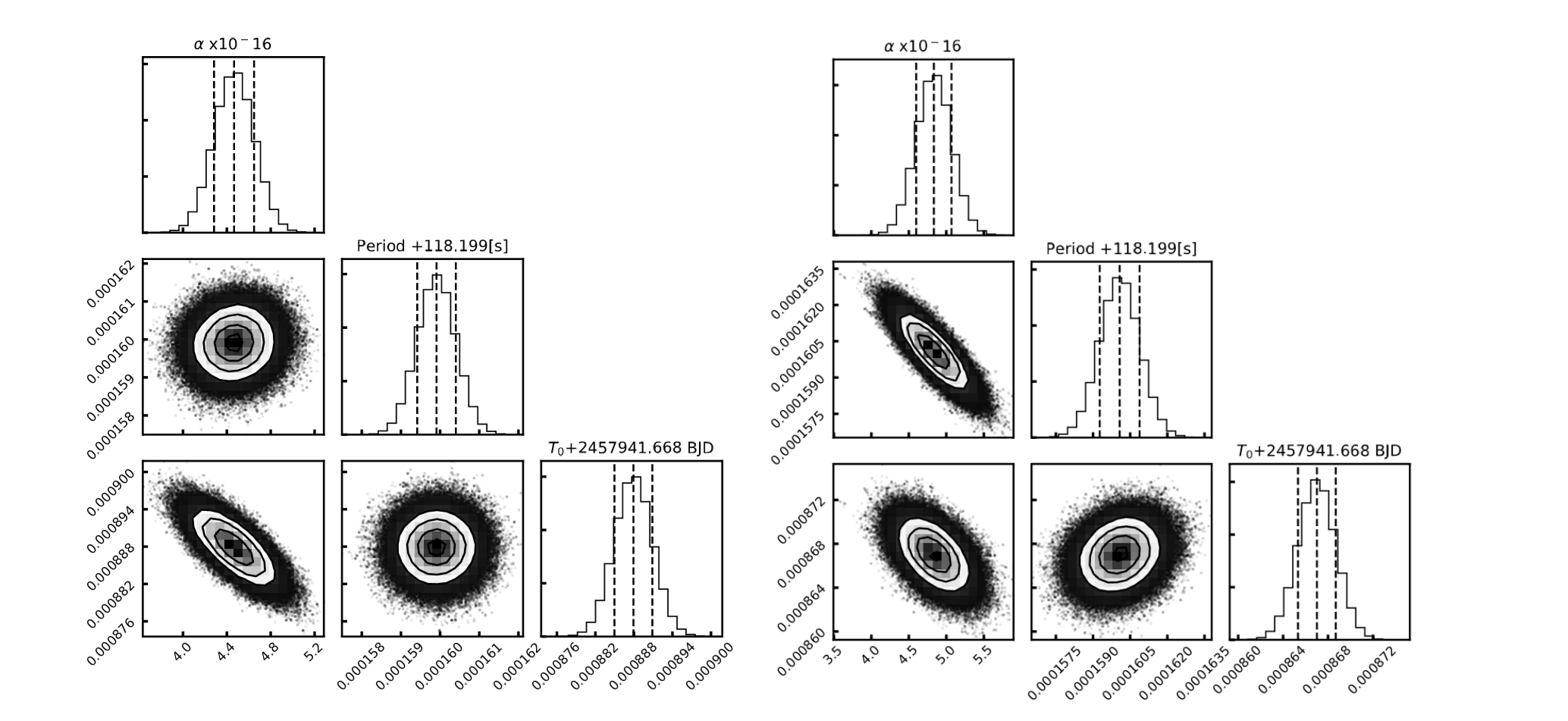}
	\caption{{\bf Left:} Corner plot of the MCMC fitting results of the SAAO+WHT-r data. {\bf Right:} Corner plot of the MCMC fitting results of the SLKT data.
	\label{corner_ALL}}
\end{figure*}

For each individual night of data we apply a linear fit to the timings and reduce the result to a single average timing for the night referenced to the first peak of that run. For the error bars of each night, we had to account for the appreciable dispersion of timings in the $O-C$ around orbital phases 0.05 and 0.55. We began by splitting the $O-C$ of the total data set into the following orbital phase bands (that match changes in dispersion as seen in Fig.~\ref{orb_correction}): 0-0.15, 0.15-0.47, 0.47-0.65, and 0.65-1. Then, we calculated the standard deviation, $\sigma_{orb}$, within each band and weighted each pulse timing by $1/\sigma_{orb}$ in the linear extrapolations back to the reference peak for that night. Thus, beat pulses occurring during `noisy' sections of the orbit have less influence on the average beat timing from an observing run. The uncertainty on the average timing for a night was calculated using the square root of the covariance matrix from the linear fit for each night of data. The SAAO+WHT-r data consist of single nights in each season, so the timings total five points with uncertainties of about $\pm 0.5$s. The SKLT data were obtained on 35 nights over four seasons, each with an average uncertainty of $\pm 0.8$s.

To estimate the beat period, we fit a first-order polynomial to the SAAO+WHT-r and the SLKT timings separately. The data sets are analyzed separately because the consistent, but uncertain, shutter offset in the SLKT timing might result in systematic error in a joint fit. The residuals to the linear fit are shown in the top panel of Figure~\ref{residuals}, and there is an obvious systematic curvature from a nonzero period derivative. The $\chi^2$ parameter from the SLKT data is very high at 542 with 33 degrees of freedom (dof).  The linear ephemeris for the SAAO-WHT-r data gives a $\chi^2 = 601$ with 3 dof. 

Fitting the data using a second order polynomial leaves no discernible trend in the residuals and the $\chi^2$ parameters are 92 (32~dof) and 0.95 (2~dof) for the SLKT and SAAO-WHT-r respectively. As found by \citet{stiller18}, the beat period is clearly not constant.

To measure the beat period change and estimate its uncertainty from this set of data, we calculate an ephemeris including a second order term in the form  \(T_{max}= T_0 + PE + \alpha E^2 \) ,
where \(T_{max}\) is the beat pulse time at maximum, \(E\) is the epoch number, \(P\) and \(T_0\) are the period and date at epoch zero, respectively. The parameter \(\alpha\) is equivalent to \(\frac{1}{2} \bar{P}_{beat} \dot{P}_{beat}\) where \(\bar{P}_{beat}\) is the average beat period and \(\dot{P}_{beat}\) is the period derivative. We apply a Markov Chain Monte Carlo (MCMC) algorithm on the data sets to determine beat ephemerides and estimate parameter uncertainties. The MCMC results (Figure~\ref{corner_ALL}) indicate that the $\alpha$ coefficents and the estimated periods for the two data sets are fairly consistent. However, $T_0$ parameters are significantly offset from each other. With the uncertainties in the final digits shown in parentheses for each parameter, the SAAO+WHT-r ephemeris is:

\begin{equation}
\begin{split} 
T_{max}(BJD) =  2457941.6688879(30) \\ 
+ 0.001368045832(06)E \\ \ \ \ 
 + 4.46(18)\times 10^{-16}E^2 ,
\end{split}
\end{equation} 

and for the SLKT data:

\begin{equation}
\begin{split}
T_{max}(BJD) = 2457941.6688670(16) \\
+ 0.001368045834(10)E \\
 +4.83(23)\times 10^{-16}E^2 .
\end{split}
\end{equation}

From the quadratic coefficient, $\alpha$, we can calculate \(\dot{P}_{beat}\) and from that the frequency derivative. Since the frequency is the inverse of the period, then the beat frequency derivative is \(\dot{\nu} = -\frac{\dot{P}}{P^2}\). 
The two frequency derivatives from the two independent data sets are:
 \[\dot{\nu}_{beat} ({\rm SAAO+WHT-r})= -4.66(19)\times 10^{-17}\  Hz\; s^{-1}\]
\[\dot{\nu}_{beat} ({\rm SLKT})= -4.98(17)\times 10^{-17}\  Hz\; s^{-1}\] 
and the other fit coefficients and their uncertainties are shown in Table~\ref{spin_values}. The last column in the table shows the difference between the SAAO+WHT-r value and the SKLT value divided by the total uncertainty. The frequency derivatives and the periods differ by about one standard deviation. The difference in the time of zero epoch are nearly six standard deviations off. We attribute this to a combination of two effects: (1) the shutter lag in the SLKT's camera, which lacks dedicated hardware to accurately measure the timestamp of each exposure, and (2) the wavelength dependence of the pulse timings, which we discuss in Sec.~\ref{sec:wavelength}.

\begin{table}
	\caption{Spin-down values comparison \label{spin_values}}
	\begin{tabular}{cccc}
	\hline
	{Value} & {SAAO+WHT-r} & {SLKT} & {$|\Delta|/\sigma^{b}$} \\
	\hline
	 $\dot{\nu}\; $[Hz s$^{-1}$] & $-4.66(19)\times10^{-17} $& $-4.98(17)\times10^{-17} $& $1.26$ \\
	P [s]& 118.19915988(52) & 118.19915918(53) & $0.943$ \\
	T$_{0}$[s] $^{a}$  & 191.58(25) & 189.99(10) & $5.91$ \\
    \hline
	\end{tabular}
	$^{\ \ a}${ +212366160000s} \\
	$^{\ \ b}${ difference in standard deviations}
\end{table}

\begin{figure}
    \centerline{
	\includegraphics[width=1.1\columnwidth]{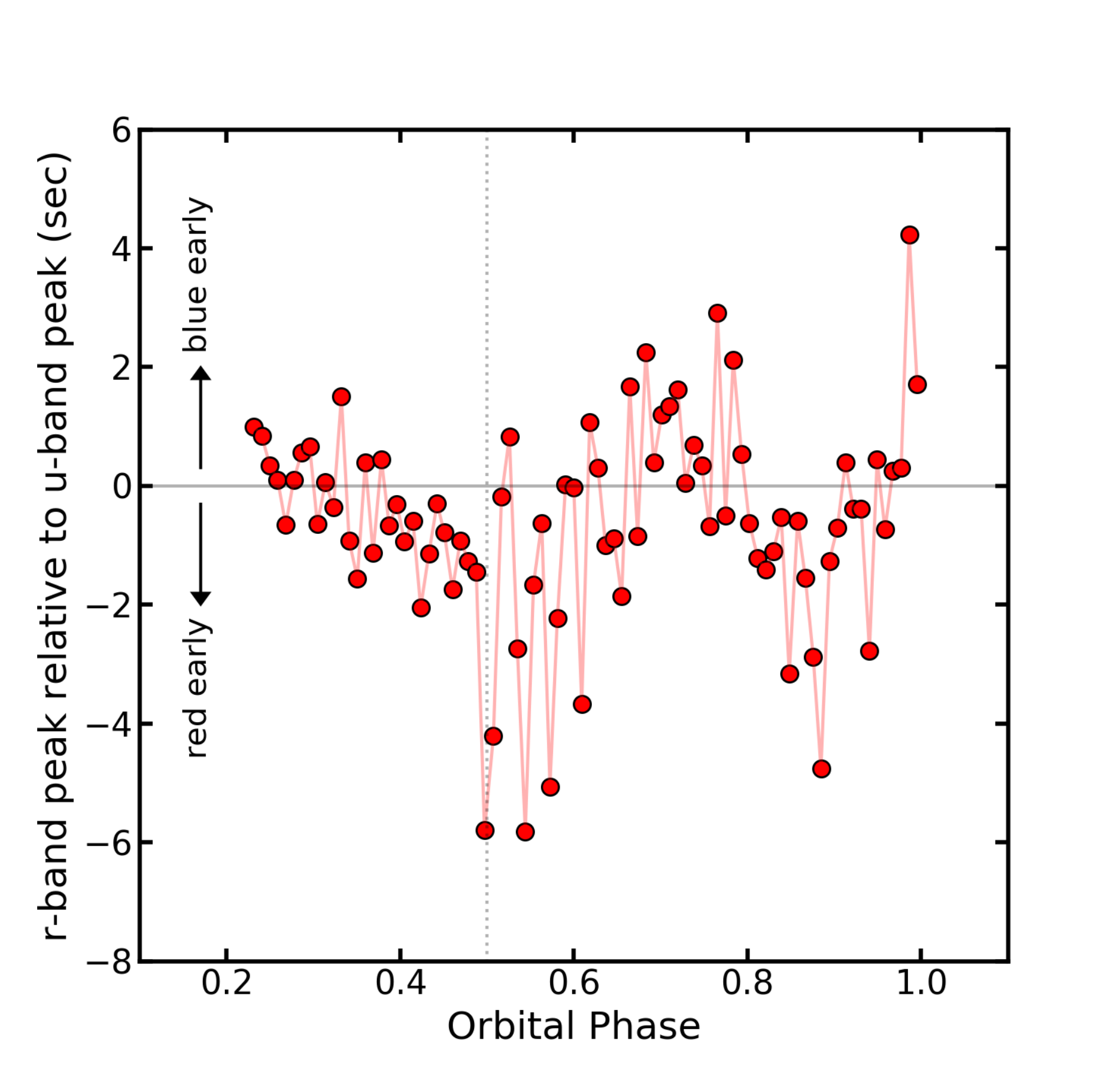}}
	\caption{Times of primary pulses in the $r$-filter relative to the times of peaks in the $u$-filter based on the ULTRACAM data. Before orbital phase 0.5 there is a small scatter in the relative peak times implying a timing precision of $\pm$0.6s. In the first half of the orbit the times of the red peaks shift from arriving a second later than the blue to nearly 2s earlier. In the second half of the orbit the relative peak times are more erratic although trends are still apparent.    \label{OC_ultra}}
\end{figure}

\begin{figure*}
	\includegraphics[width=0.9\textwidth]{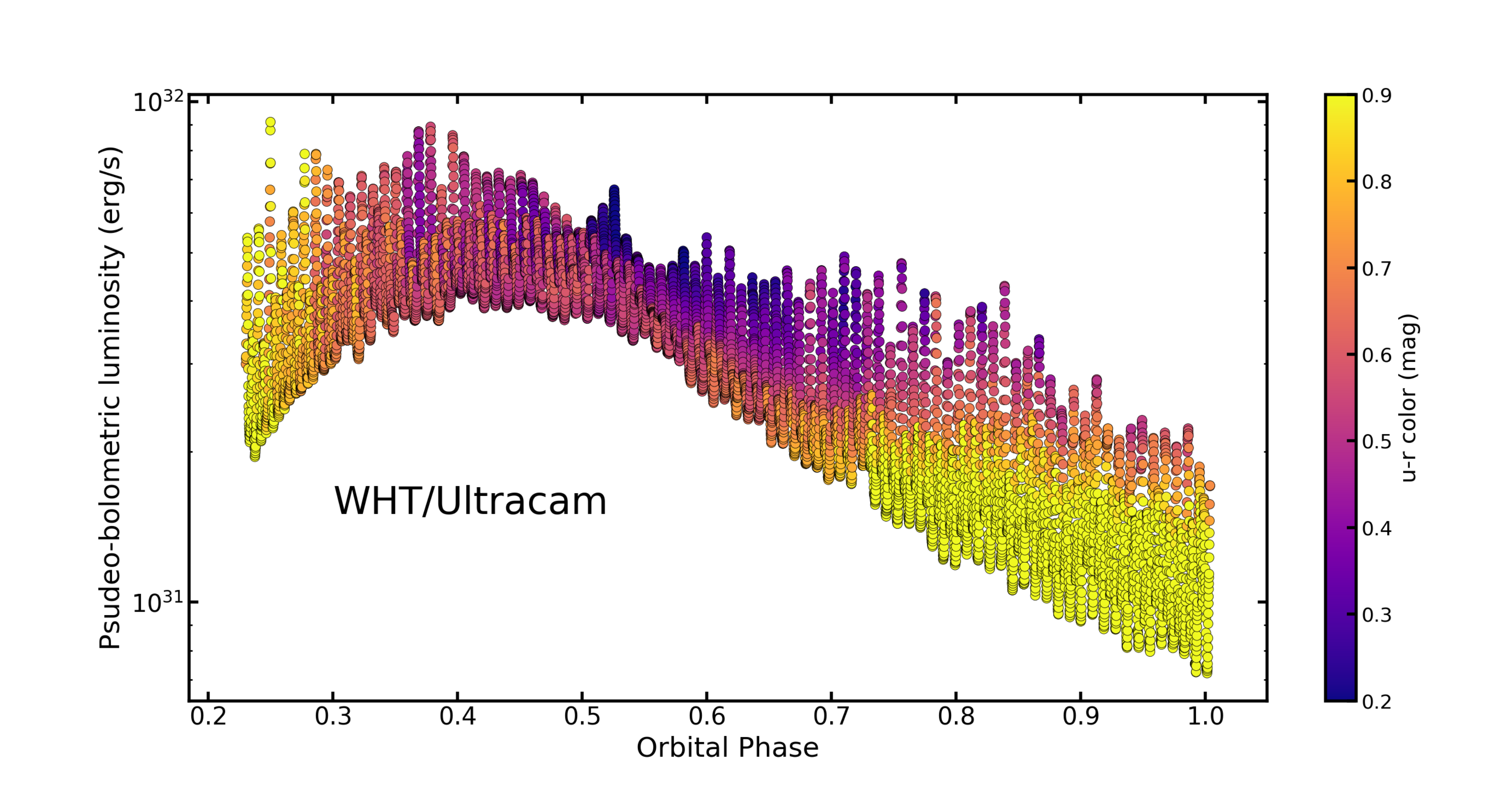}
	\caption{The pseudo-bolometric light curve of AR~Sco constructed from the ULTRACam photometry. The $u-r$ color of each observation is indicated by the color of the symbol.
	\label{bolometric}}
\end{figure*}

\subsection{Systematic Uncertainties in the Spin-Down Rate}

\subsubsection{Pulse Times vs Wavelength} \label{sec:wavelength}

Combining observations of AR~Sco from several telescopes, cameras, and filter systems may lead to systematic pulse timing errors. For example, \citet{stiller18} found camera shutter-lag differences between data sets that generate $O-C$ timing offsets. If the pulse shapes are color-dependent, then light curves taken through different filters will show timing shifts that are not related to the WD spin-down.

The multi-filter imaging obtained by ULTRACAM on the WHT provides an excellent test of the dependence of pulse peak times on wavelength. Photometry of AR~Sco was obtained simultaneously in SDSS $u$, $g$, and $r$ filters, making it possible to estimate a pulse-by-pulse timing shift between filters. We focus on $r$ and $u$-bands, the two filters with the widest span of wavelength. As shown in Figure~\ref{lightcurve}, the primary pulses are clearly present in both filters, with the main differences being in the pulse amplitudes. Secondary pulses can be less defined than the primary peaks and can even disappear at some orbital phases for the $u$-band data.

Starting from the $r$-band primary pulse timings derived for the beat ephemeris analysis, we fit the peaks for the $u$-band photometry and plot the difference in peak times between the two filters in Figure~\ref{OC_ultra} as a function of orbital phase. Note that the ULTRACAM data from 2005 covered only 80\%\ of an orbit. There are clearly systematic pulse timing differences between the two filters, and these show a complicated relation with orbital phase. Between orbital phases 0.2 and 0.5, the difference in pulse times between the $r$ and $u$-bands shows a very small scatter and a clear trend. At first, the pulse arrives earliest at blue wavelengths, but by orbital phase 0.5 the pulse arrives first in the red filter. A linear fit to the trend shows that the difference in peak times between these filters amounts to 3.5~s over the first half of the orbit, or 3\%\ of the beat period. A similar wavelength dependent phase shift has been seen by \citet{marsh19}.

Taking into account the linear trend, the standard deviation in the relative timings over the first half the orbit is 0.64~s. This scatter is small because we are looking at the same pulses over the same time windows and the only variable is the wavelength. 

The scatter in the relative pulse times between filters becomes large after orbital phase 0.5. This jump in the timing scatter amplitude after phase 0.5 is seen even in a single filter and reflects a fundamental change in light curve properties at superior conjunction. Between orbital phases 0.5 and 0.65, the red pulse can arrive as much as 6~s ahead of the blue, while at other times they peak simultaneously. Over the second half of the orbit, the relative times of the red and blue pulses oscillate between $\pm 2$s.

The cause of these color variations is certainly not clear, but their existence may be a clue to the nature of the beat and spin pulses. For example, if the beat and spin pulses are adding together over the first half of the orbit, then the higher frequency spin pulses would switch from arriving behind the beat pulse to ahead. So the color timings in the first half can be explained as the spin pulse being redder than the beat pulse.

To avoid systematic errors in the beat pulse times, one should be consistent in the filters used for the pulse timings. Timing differences of seconds between the $r$ and $u$ filters are seen over certain orbital phases, and errors of this size would have a major impact on the estimate of $\dot P$. The SAAO data were obtained with a red-sensitive detector, and that was matched with r-filter data from WHT for consistency. The SLKT photometry was all taken with an unfiltered CCD. So our timing analysis minimizes any systematic color dependence on pulse times.

\subsubsection{Ruling out a proper-motion contribution to $\dot P$}

Previous studies of the spin-down have not considered the possibility that some fraction of the observed $\dot{P}$ may be attributable to AR~Sco's proper motion (PM). The transverse component of the relative space motion of a star produces a time-varying radial velocity, and this varying Doppler shift generates an apparent period derivative. \citet{pulsar70} applied this effect to period changes in the newly discovered pulsars. More recently, \citet{pajdosz95} discussed this anomalous $\dot P$ in the context of rapidly pulsating ZZ~Ceti stars, but it applies to rotational periods as well. Critically, a star's proper motion will always induce a positive $\dot{P}$ term in studies of a periodic signal \citep{pajdosz95}---making it important to examine whether AR~Sco's much-heralded positive $\dot{P}$ might be attributed to this effect.

Using Eq.~8 in \citet{pajdosz95} and Gaia DR2 astrometry, we find that AR~Sco's proper motion induces a unitless beat period derivative of $\dot{P}_{pm, beat} = 9 \times 10^{-17}$. This is four orders of magnitude smaller than the observed $\dot{P}_{beat}$ and, therefore, it contributes an insignificant error to our measurements of the WD's spin-down. Likewise, when calculated for the orbital period, $\dot{P}_{pm, orb}$ is almost 700 times too small to impact the constraints on the orbital-period derivative from \citet{peterson}.


Although this effect is far too small to measurably influence AR Sco's $\dot{P}$, it should be taken into account should additional AR~Sco-type objects be discovered with apparent period derivatives.

\subsection{Pseudo-Bolometric Flux}

The ULTRACam photometry from 2015 obtained data in the $SDSS-r$, $SDSS-g$, and $SDSS-u$ simultaneously and this permits an analysis of color variations and pseduo-bolometric flux estimates. We convert the uncalibrated flux measurements from the ULTRACam CCDs to magnitudes. Here, we use the average spectrum of AR~Sco covering nearly an orbit \citep{garnavich19} and assume the average flux in each filter matches the corresponding average flux densities in the spectra. The resulting light curves are shown in Figure~\ref{lightcurve}, along with a detailed comparison between the $SDSS-r$ and $SDSS-u$ filters.

There are striking differences between the red and ultraviolet (UV) light curves of AR~Sco. For example,
over the second half of the orbit, the amplitude of the primary UV pulses are larger than the amplitude at red wavelengths. \citet{stiller18} and \citet{potter18a} showed that in unfiltered (dominated by red wavelengths) photometry the pulse amplitude peaks at orbital phase 0.25 with a weaker maximum around phase 0.8. At UV wavelengths, this is reversed and the strongest peaks occur at the late orbital phases (Figure~\ref{lightcurve}).

\begin{figure}
    \centerline{
	\includegraphics[width=1.1\columnwidth]{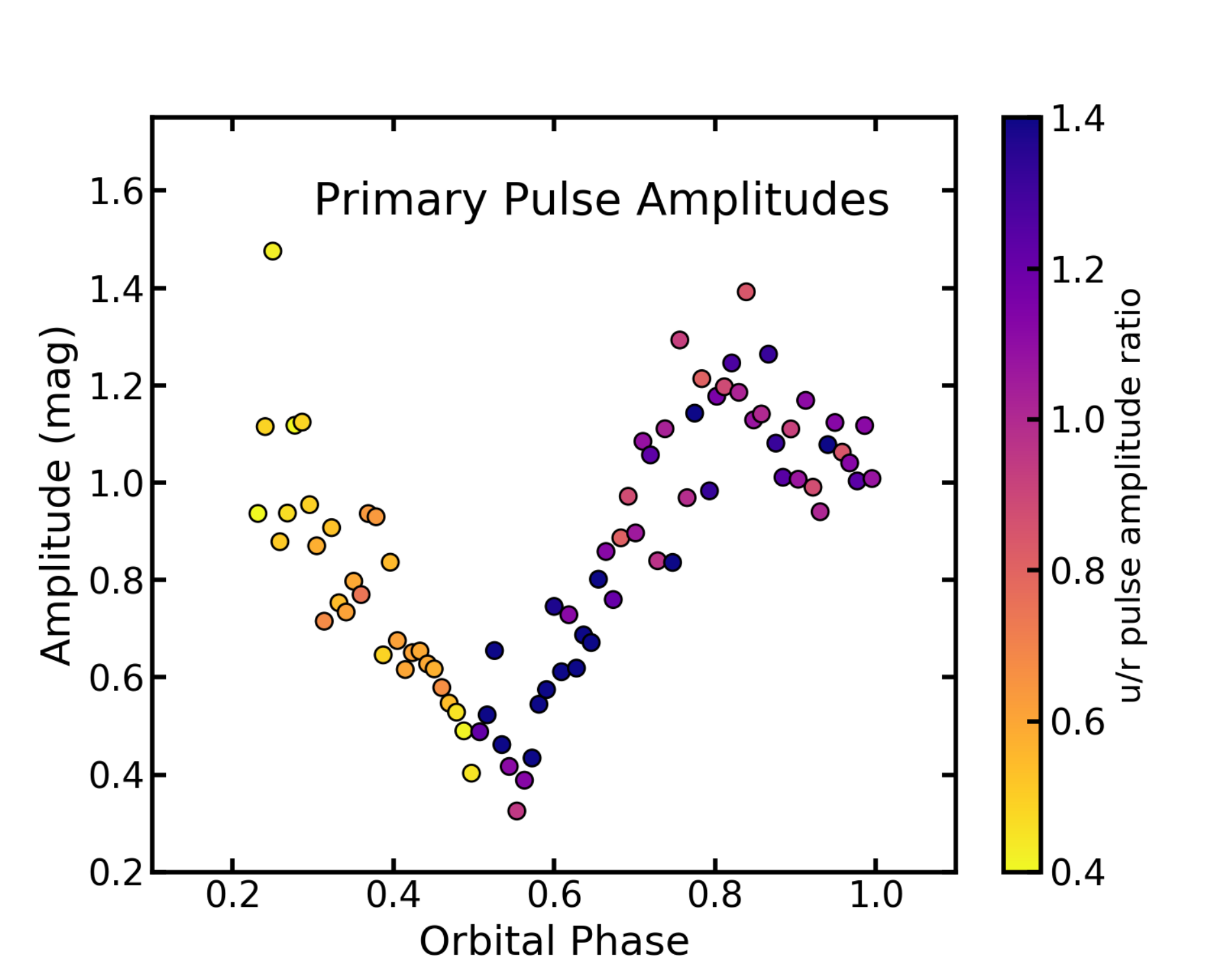}}
	\caption{The primary pulse amplitude from the pseudo-bolometric luminosity. The amplitude is the magnitude difference between the minimum and maximum luminosities for each pulse. The amplitude ratio between the u-band and r-band fluxes is shown as a color for each point. Starting at phase 0.50, the pulse amplitudes are dominated by blue wavelengths when compared with the first half of the orbit. Thus, the bolometric pulse amplitudes are fairly symmetric about superior conjunction, in contrast to light curves obtained at red wavelengths. \label{amplitude}}
\end{figure}

\begin{figure*}
    \centerline{
	\includegraphics[scale=0.6]{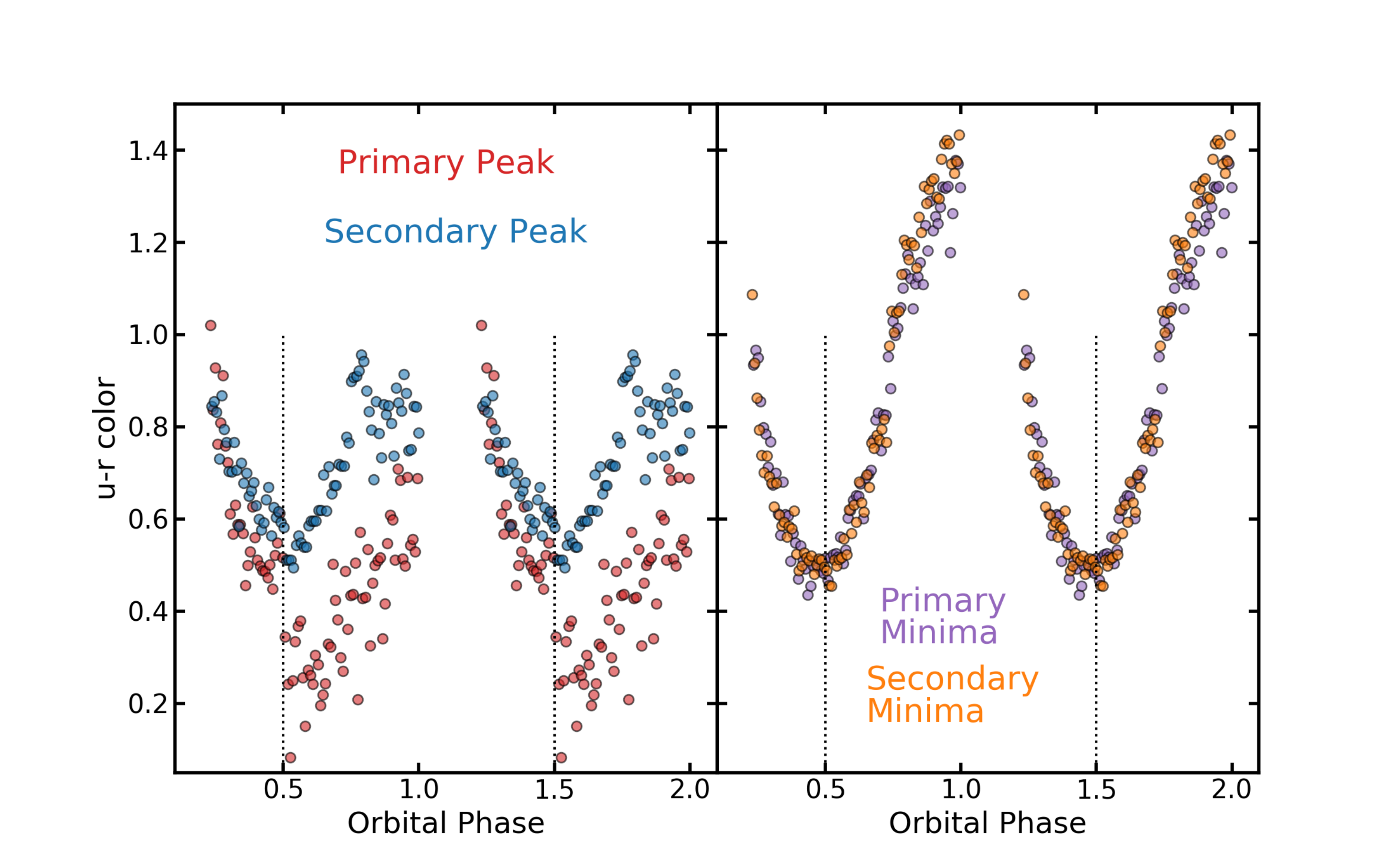}}
	\caption{{\bf left:} The $u-r$ color index at the peak flux for the primary and secondary pulses. Over the first half of the orbit the colors of the primary and secondary peaks are very similar, but the primary peak suddenly turns blue at orbital phase 0.5. {\bf right:} The $u-r$ colors of the minima over each beat cycle. The primary minimum is defined as the dip after the primary pulse. The color index has a large range over the orbit. The primary and secondary minima have similar colors over the dataset. \label{color}}
\end{figure*}

The secondary pulses also show significant differences in amplitude between the red and UV bands. In the red, the primary pulses weaken after orbital phase 0.5 and reach a comparable amplitude as the secondary pulses. But in the UV, the primary pulses remain quite robust and it is the secondary pulses that weaken to the point of being undetectable. Clearly, the primary peaks will have $u-r$ colors that are quite blue while the secondary pulses will become more red.

The ULTRACam high-cadence multi-color photometry allows an estimate of the total flux from AR~Sco over a broad wavelength range. The $SDSS-u$, $SDSS-g$, and $SDSS-r$ bands are centered at 3557~\AA, 4825~\AA, and 6261~\AA\ respectively \citep{dhillon07}. We convert the magnitudes in each filter to their corresponding flux densities and fit a quadratic polynomial at each set of exposures. The $SDSS-u$ band images had twice the exposure time as the other filters so their magnitudes were interpolated to the same cadence as $SDSS-r$. The quadratic fit was then integrated between 2500~\AA\ and 7000~\AA\ to give the pseudo-bolometric flux. Assuming a distance of 117.8~pc, we convert to a luminosity ($erg\; s^{-1}$). The resulting light curve is shown in Fig~\ref{bolometric}.

The bolometric light curve suggests that the primary pulses are more uniform in amplitude around an orbit than is seen at red wavelengths. This is clearly demonstrated in Figure~\ref{amplitude} where the difference between a pulse maximum and its nearest minimum is displayed. The weakest pulse amplitudes are seen soon after orbital phase 0.5. There is a short spike in blue-dominated pulses at orbital phase 0.52, but it is not clear from this one orbit if that repeats every cycle. The color of the points indicates the relative pulse amplitude in the $u$ and $r$ bands. The pulses are dominated by red emission during the first half of the orbit, but after orbital phase 0.5 the primary pulses are strongest at blue wavelengths. The sharp transition in color at binary conjunction is clearly seen in Figure~\ref{amplitude}.

The variation in colors of the primary and secondary peaks are shown in Figure~\ref{color}. The $u-r$ color of the secondary pulses reach a minimum (most blue) at orbital phase 0.51 and follow a parabolic trajectory between orbital phases 0.2 and 0.8. The $u-r$ color of the primary pulse peaks are very similar to that of the secondary peaks over the first half of the orbit, but are significantly more blue over the second half. As noted above, there appears a discontinuity in the colors at orbital phase 0.5 where the peak colors suddenly become 0.3~mag more blue than the secondary peaks. Figure~\ref{color} also indicates that the scatter in the primary peak colors shows a marked increase beyond orbital phase 0.5. 

The color of the minima between pulses is very well-behaved (Figure~\ref{color}). The primary minimum is defined as the faintest bolometric flux following a primary peak and the secondary minimum follows directly after a secondary peak. Both primary and secondary minima follow a parabolic-like color variation over an orbit with the bluest color occuring just before orbital phase 0.5. The $u-r$ color gets extremely red near orbital phase 1.0, and the primary minima appear marginally more blue than the secondary minima between phases 0.8 and 1.0.

The color discontinuity in pulse peaks at orbital phase 0.5 suggests that the viewing angle geometry plays an important role in light curve of AR~Sco. Orbital phase 0.5 is at superior conjunction when the secondary star is precisely on the far side of the WD as seen from the Earth. This discontinuity at superior conjunction is also seen in the peak timings as shown in Figure~\ref{OC_ultra}. The color and timing discontinuities suggest that a pulse component comes into, or disappears from, our view near conjunction. The strengthening of a blue component immediately after syzygy would explain the reduced color indices of the peaks in the second half of the orbit, as well as the rapid change in peak times as seen in Figure~\ref{OC_ultra}. 

In the model developed by \citet{potter18b}, the beat pulses consist of synchrotron radiation that originates in the outer magnetosphere as relativistic electrons accelerate along field lines towards magnetic mirror points. Although the resulting synchrotron emission is fixed within the WD's rotational rest frame, it is enhanced whenever one of the WD's magnetic poles sweeps past the secondary, receiving an injection of relativistic electrons. Due to a combination of our changing viewing angle and synchrotron beaming effects, the amplitude of the beat pulses is largest near orbital phases $\sim0.3$ and $\sim0.8$. One of the most interesting predictions of this model is that near superior conjunction, the magnetic pole responsible for the primary beat pulse changes---\textit{i.e.,} the observed synchrotron radiation comes from the opposite magnetic hemisphere. The abrupt pole-switch near superior conjunction predicted by their model could therefore explain why there is such a sharp discontinuity in the pulse color at the same orbital phase. However, although we might also anticipate a simultaneous switch to redder colors for the secondary pulses, no such jump in color is seen.

\section{Conclusion}

We have measured beat frequency derivative of
\(-4.82(18)\times10^{-17} Hz\; s^{-1}\), by combining five years of data from SAAO, WHT and the SLKT. This is consistent with the \(-5.14(32)\times 10^{-17} Hz\; s^{-1}\) estimated by \cite{stiller18}, but our estimate has nearly twice the precision. Constraints on the change in the binary orbital frequency \citep{peterson} imply that the change in the beat is identical to the spin-down frequency of the WD, which is now detected at more than 25$\sigma$. 

We find that the timing of the beat pulses is wavelength dependent, with variations of more than $\pm 2$s in peak pulse arrival times between UV and red bands as a function of orbital phase. The wavelength dependence on timing may induce systematic errors in future ephemeris determinations unless care is taken in obtaining consistent data. This color dependence on pulse arrivals may also be useful in constraining the emission mechanism and geometry in AR~Sco. 
The pulsed emission from AR~Sco shows systematic color variations over an orbit. The primary pulses tend to be redder before superior conjunction when compared to pulses after conjunction. The change in color is discontinuous at orbital phase 0.5. In contrast, the secondary pulses show a continuous color variation with the bluest peaks occurring at superior conjunction. This jump in the colors of the primary pulses as the binary passes superior conjunction suggests that our view of the emission source undergoes a discontinuous change as may happen by viewing a switch in the magnetic pole dominating the emission.


\section*{Acknowledgments}

We are grateful to T. Marsh for providing the WHT photometry used here. We thank R. Stiller for obtaining some of the SLKT observations. DAHB \& SBP acknowledge their research support from the National Research Foundation of South Africa.



\bsp	
\label{lastpage}
\end{document}